\documentclass[journal=jacsat,manuscript=article]{achemso}

\usepackage{graphicx, caption}
\usepackage{xcolor}
\usepackage{adjustbox}
\usepackage{tabularx}
\usepackage{mathtools}
\usepackage[version=3]{mhchem} 

\usepackage{amssymb}
\usepackage{braket}
\usepackage{esvect}
\usepackage{listings}
\usepackage{mathptmx}
\usepackage{amsmath}

\title{A Physically-informed Graph-based Order Parameter for the Universal Characterization of Atomic Structures}
\author{James Chapman}
\affiliation[Lawrence Livermore National Laboratory]{Physical and Life Sciences Directorate,  Lawrence Livermore National Laboratory, Livermore, CA}
\email{chapman37@llnl.gov}

\author{Nir Goldman}
\affiliation[Lawrence Livermore National Laboratory]{Physical and Life Sciences Directorate,  Lawrence Livermore National Laboratory, Livermore, CA}
\alsoaffiliation{Department of Chemical Engineering, University of California, Davis, California 95616, United States}

\author{Brandon Wood}
\affiliation[Lawrence Livermore National Laboratory]{Physical and Life Sciences Directorate,  Lawrence Livermore National Laboratory, Livermore, CA}

\begin{document}
\maketitle

\begin{abstract}

A new graph-based order parameter is introduced for the characterization of atomistic structures. The order parameter is universal to any material/chemical system, and is transferable to all structural geometries. Three sets of data are used to validate both the generalizability and accuracy of the algorithm: (1) liquid lithium configurations spanning up to 300 GPa, (2) condensed phases of carbon along with nanotubes and buckyballs at ambient and high temperature, and (3) a diverse set of aluminum configurations including surfaces, compressed and expanded lattices, point defects, grain boundaries, liquids, nanoparticles, all at non-zero temperatures. The aluminum configurations are also compared to existing characterization methods for both speed and accuracy. Our order parameter uniquely classifies every configuration and outperforms all crystalline order parameters studied here, opening the door for its use in a multitude of complex application spaces that can require fine configurational characterization of materials.
  
\end{abstract}

\section{Introduction}

The structure-property relationship forms the basis of modern materials science/chemistry, as the discovery of new materials and further understanding of existing ones neccesitates the need to probe the atomic/nano regime \cite{doi:10.1063/1.3548889,Schmidt2019,archer_order_2014,Zuo2020,D1CP01349A}. At the heart of this relationship is the fundamental principle that the arrangement of atoms dictates the behavior of the material throughout a spectrum of length and time scales. Reliably capturing structure-property relationships plays a vital role in fields such as crystal structure prediction \cite{Fischer2006}, the dynamic evolution of complex defect networks \cite{10.3389/fmats.2017.00034}, and the construction of interatomic potentials \cite{doi:10.1063/5.0009491,doi:10.1021/acs.jpcc.9b03925,PhysRevB.100.024112} among others. Employing order parameters that can uniquely capture local atomic geometries are necessary to adequately characterize phase transitions from molecular dynamics simulations \cite{doi:10.1063/1.3656762,PhysRevLett.47.1297}, and also plays a critical role in free energy calculations \cite{doi:10.1080/00268979909483070,Eslami2017}. Therefore, characterizing the atomic structure of a material in a computationally efficient and physically meaningful way is critical in understanding the material's underlying properties. 

However, this characterization is often non-trivial, especially for disordered systems in which the underlying symmetry of the atomic geometries is difficult to determine \cite{PhysRevB.51.5768,Tian_2011}. Throughout the decades, many schemes have been proposed to capture various portions of this ordered-disordered spectrum such as the Common Neighbor Analysis (CNA) \cite{Honeycutt1987}, Adaptive CNA (A-CNA) \cite{Stukowski_2012}, Centrosymmetry parameter (CNP) analsysis \cite{PhysRevB.58.11085}, Voronoi analysis \cite{noauthor_druckfehlerverzeichnis_1908}, Bond Order Analysis such as the Steinhardt Order Parameter (SP) \cite{doi:10.1080/01418618108235816}, and the Bond Angle Analysis (BAA) \cite{PhysRevB.73.054104}. Each method has shown varying degrees of success, with each scheme playing a vital role in capturing specific classes of materials phases \cite{Stukowski_2012}. However, to our knowledge none of the above methodologies can adequately characterize all phases of material in a unique and physically meaningful manner. Voronoi, SP and other bond-order algorithms generally fail to capture the differences in crystalline systems with compressed and/or expanded lattices as well as those experiencing atomic perturbations close to the melting temperature of the material phase \cite{KEYS20116438}. While methods such as CNA and A-CNA overcome these pitfalls with a more robust underlying algorithm, they ultimately break down in situations where in the material symmetry is lost or difficult to comprehend \cite{DENG2018195}. In fact, all of the above algorithms struggle to capture the subtle differences in the local coordination environment when the underlying geometric symmetry is either broken or exists only at short-range such as the environments encountered in grain boundaries, surfaces, liquids and amorphous structures \cite{10.3389/fmats.2019.00120}. 

More mathematically involved methods, such as the Smooth Overlap of Atomic Positions (SOAP) \cite{C6CP00415F}, the Behler-Parinnello symmetry functions (BP) \cite{doi:10.1063/1.3553717}, and the Adaptive Generalizable Neighbhood Informed features (AGNI) \cite{doi:10.1021/acs.jpcc.9b03925}, rely on sophisticated functional forms with a plethora of tunable parameters to map an atom's local environment to an invariant mathematical space. While these methods are often accurate \cite{Chapman2020,Deringer2018,PhysRevB.95.094203,Rosenbrock2017,doi:10.1063/1.4712397}, they are also computationally cumbersome \cite{Zuo2020} and require the manual tuning of their corresponding parameter sets for every new material studied. Methods such a convolutional neural networks (CNN) \cite{duvenaud2015convolutional,Kearnes2016}, graph neural networks (GNN) \cite{doi:10.1021/acs.chemmater.9b01294,zeng2018graph}, and variational autoencoders (VEA)\cite{Batra2020} can alleviate both the cost and manual parameter fitting of SOAP, BP, and AGNI, but require large amounts of reference data to train the models. In particular, these methods can be difficult to train for materials with complex chemical phase spaces (e.g., detonations of energetic materials\cite{Lindsey_DNTF}). This can hinder both the generalizability and transferability of these models to new configurations which are not previously characterized within the training set. Graph theoretical methods such as those employed in MoleculaRnetworks \cite{https://doi.org/10.1002/jcc.22917} and ChemNetworks \cite{https://doi.org/10.1002/jcc.23506} have been used to analyze small molecules with good success. However, such methods rely on properties of the graph representations that are not unique, such as the geodesic distance of the graph, and are thus not suitable for broad material classes that have similar structural properties. This includes oxides, metals, ceramics, and/or physical conditions such as extreme pressures, grain boundaries, surfaces and nanoparticles.

In this work, we overcome these challenges through development of a physically intuitive and computationally efficient framework, henceforth referred to as the Scalar Graph Order Parameter (SGOP).  Our approach uses a semi-empirical graph isomorphism metric to not only characterize the complexity found across a material's phase space, but also to alleviate the pitfalls and bottlenecks of the aforementioned methods. We also discuss a Vector Graph Order Parameter (VGOP), which allows for linear combinations of different SGOP values in order to add a high degree of sensitivity to our analysis. In general, these order parameter characterize the underlying graph of the network contained within a configurations of atoms. This characterization can be broken down into three parts: (1) identification of subgraphs contained within the system, (2) determination of the shape of the subgraphs, which is motivated to resemble the entropy of the subgraph, and (3) calculations of the connectivity of the subgraphs, which is determined via the subgraph's degree matrix. Unlike other methods, where the characterization is performed within some high-dimensional mathematical space meant to represent atomic environments, or in a space that is too simplistic to distinguish between subtle structural differences, our method aims to differentiate between the graphs that represent atomic structures, specifically. It is this difference that allows for not only a reduction in the cost and complexity of the algorithm, when compared to existing methodologies, but also provides a more physically intuitive understanding into the relationship between configurations, based on the resulting order parameter value.

We demonstrate the use of SGOP and alternately VGOP for three test cases: (1) a wide variety of liquid lithium phases under extreme pressures, (2) a diverse set of carbon structures at various temperatures \cite{doi:10.1021/acs.jpca.0c07458}, and (3) a set of aluminum data, derived via ab initio molecular dynamics, which spans vast regions of its phase space \cite{Batra2020}. For the case of liquid lithium we showcase that a single user-defined parameter within SGOP yields an order parameter that can correctly classify all phases in a physically intuitive manner. We next demonstrate the use of VGOP to uniquely characterize not only bulk phases of carbon, but also more complex geometries such as nanoparticles and nanotubes. We conclude our manuscript by illustrating our method's ability to distinguish between a multitude of environments of aluminum including compressed and expanded lattices, point defects such as vacancies and di-vacancies, planar defects such as grain boundaries and surfaces, as well as various nanoparticles. The exactness and efficiency of the proposed methodology allows one to reliably characterize the structural subtleties between materials in a computationally efficient and physically informed manner.

\section{Results}
\subsection{Lithium Liquids}

Our first evaluations of the SGOP framework, which is discussed within the computational details section, revolves around the characterization and classification of a multitude of liquid lithium phases under extreme pressures ($P \in [30GPa, 350GPa]$). Previous works have indicated that the configuration space of the liquid phases spans a vast domain, with each liquid phase showing structural differences when compared to results from a different pressure\cite{Guillaume2011}. These structural dissimilarities result in strong differences in properties such as the vibrational density of states, which ultimately govern the self-transport behavior of the material \cite{PhysRevLett.108.055501}. Previous density functional theory calculations have shown that, within a given temperature range, there is a strong linear correlation between the self-diffusion constant and the density of the liquid phase \cite{PhysRevLett.108.055501}. The coupling of these two properties allows one to make predictions on unknown phases at high pressures without the need for performing non-trivial and expensive simulations and/or experiments.

\begin{figure*}
        \centering
    	\includegraphics[width=1.0\textwidth]{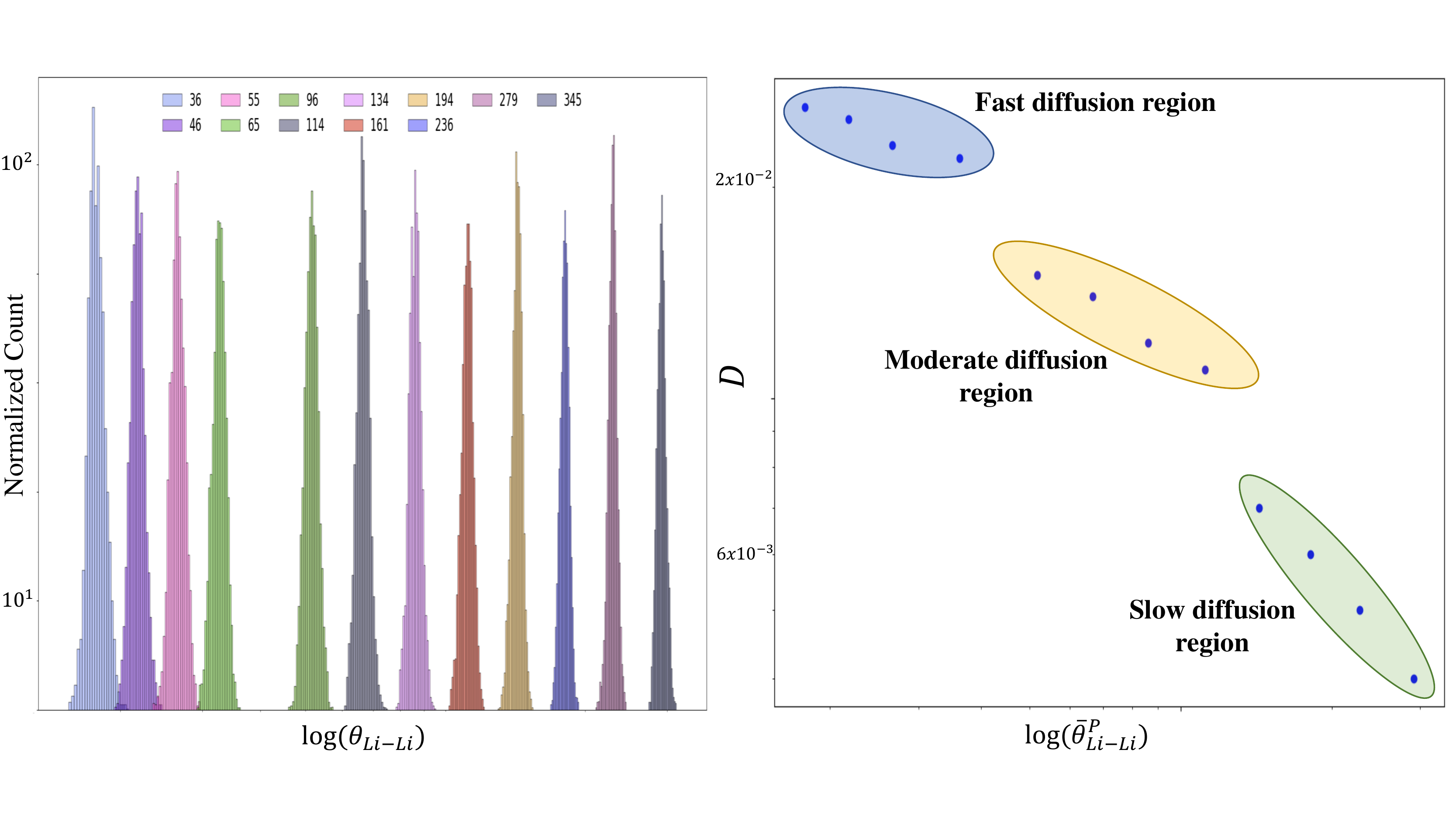}
    	\caption{(a) Histograms representing the distribution in SGOP values for each liquid phase of lithium. Colors represent the different phases, and are defined in the plot by the external pressure on the simulation box obtained from DFT. (b) Self diffusion constant values, obtained from the mean square displacement of the ab initio molecular dynamics trajectories plotted as a function of the mean of each histogram, shown in (a). Three diffusion regions are highlighted, and are determined by the underlying structure of the vibrational density of states of each trajectory, calculated from the velocity auto-correlation function. $\theta^{P}$ is defined as the SGOP for a specific phase $P$.}
    	\label{fig:Li}
\end{figure*}

Figure \ref{fig:Li} highlights the ability of a single SGOP value to characterize the complexity of the lithium liquid phase space. The SGOP values shown here were calculated from a  Graph Coordination Network (GCN), which is discussed within the computational details section, which employed a $R_{c}$ of 2.5 $\AA$. Figure \ref{fig:Li} (a) shows histograms of the SGOP values for each liquid phase. One can clearly see the separation of each phase, indicating that the SGOP values are capable of characterizing the unique differences in local geometry encountered within each phase, but also the spread within a specific phase. The structures encountered within a liquid phase, over some period of time, will oscillate about an equillibrium point, assuming that all external conditions are held constant. Figure \ref{fig:Li} (a) provides a visual representation of these perturbations, with the width of each normal distribution representing the extent of the spread for a given phase.

Figure \ref{fig:Li} (b) showcases the SGOP's ability to reproduce the underlying trend of density versus self-diffusion constant. Diffusion constant values were calculated from the mean square displacement for each liquid phase using a simple Fickian diffusion model \cite{Berthier_2005}. Figure 1b tracks the changes in the self-diffusion constant as a function of the mean SGOP value from each phase's histogram, shown in Figure \ref{fig:Li} (a). From this relationship we can correctly identify three diffusion regions: (1) fast diffusion occurs in low-density phases, (2) moderate diffusion occurs in phases that are more dense than the low-density regime, but do not exhibit ``crystal-like'' properties, and (3) slow diffusion occurs in highly-compressed phases which behave more closely to a crystal phase than a liquid one. The SGOP histogram averages are able to classify not only the structures within each liquid phase, but also correctly identify unique self-diffusion regions across a vast configuration space. This clearly indicates that one can use the SGOP values as inputs to predictive models.

\subsection{Carbon}

While the structures encountered within the liquid lithium phase space are highly complex, they required only a single SGOP to classify the phases. This was in part due to the density acting as a the sole property needed to characterize the local coordination environment. However, as one aims to characterize the multitude of unique structural motifs within a material's phase space, a single SGOP may not be unique enough to differentiate between local atomic geometries. One example of this is elemental carbon, which exhibits a rich configuration space that includes both two-dimensional and three-dimensional structures, nanotubes containing varying amounts of free-volume, and nanoparticles that exist in many shapes and sizes. This diversity of structures and coordination numbers readily indicates the need for multiple SGOPs to adequately represent various portions of the coordination environment. 

Here we use an extension of the SGOP formalism, called the Vector Graph Order Parameter (VGOP), which is discussed within the computational details section, to characterize a previously created and highly diverse carbon dataset \cite{delRio2020}. Due to the presence of different phases with subtle structural differences, such as graphite vs. diamond, a $R_{c}$ set of $(3 $\AA$, 4 $\AA$, 5  $\AA$, 6  $\AA$)$ was chosen, which was determined via the peaks the radial distribution functions from each material in the data set. The resulting VGOP can be thought of as a feature set, similar to those discussed earlier, but with the significant advantage of both small size and easy physical interpretability. As discussed in the computational details section, each VGOP was normalized and decomposed using PCA. Information regarding the PCA metrics can be found in the supplemental information. 

\begin{figure*}
        \centering
    	\includegraphics[width=1.0\textwidth]{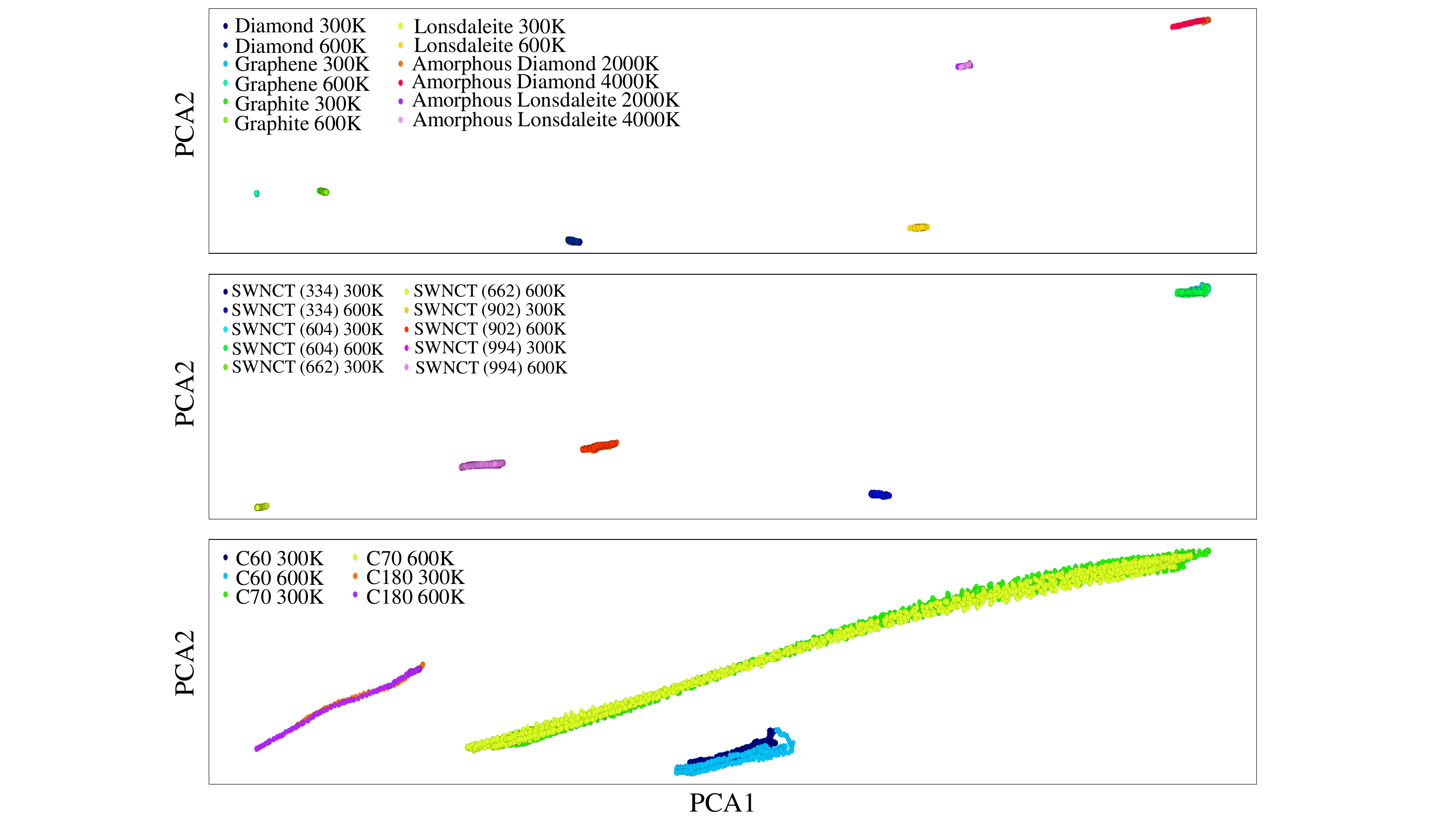}
    	\caption{Principal component analysis plots, calculated from the VGOPs of the carbon dataset's GCNs, indicating the unsupervised clustering of a multitude of structural motifs: (a) the bulk phases of carbon (graphene, graphite, diamond, and lonsdaleite), (b) single-walled carbon nanotubes of varying length and radius, and (c) buckyballs of varying diameter. }
    	\label{fig:C}
\end{figure*}

Figure \ref{fig:C} (a) indicates the VGOP's ability to characterize the various ``bulk'' phases of elemental carbon. All phases (graphene, graphite, diamond, and lonsdaleite) are clearly differentiated, and perhaps more importantly, are clustered in a physically intuitive manner. Graphene is clustered near graphite, but far from both diamond and lonsdaleite. Graphite is clustered between graphene and diamond, while lonsdaleite (hexagonal diamond) finds itself clustered near diamond but far from both graphene and graphite. Amorphous diamond and lonsdaleite also cluster near one another, but are located in a unique portion of the PCA space, when compared to the ordered crystal phases. As was the case with lithium, the VGOP not only characterizes each phase correctly, but also classifies them in a physically informed manner, providing the user with an intuitive interpretation pathway. One important aspect of these results can be seen in the classification of the amorphous configurations. The VGOP correctly classifies the structures encountered during each trajectory as similar despite the large disparity in the temperature used to generate each amorphous phase (2000 vs. 4000~K). The high fidelity of the VGOP framework can allow for precise analysis of phenomena such as phase transitions and/or for free energy calculations, where an easy and clear distinction between material phases is vital.

Similar clustering trends exist in Figure \ref{fig:C} (b) and \ref{fig:C} (c) for nanotubes and buckyball-like nanoparticles respectively. In Figure \ref{fig:C} (b), nanotubes with small radii are isolated from those with large radii in the PCA space. This again makes intuitive sense, as the local atomic coordination environment will change as a function of the nanotube's radius. This can also be seen in Figure \ref{fig:C} (c) for the case of small buckyball-like nanoparticles where particles with a fewer number of atoms that are more densely packed than those with a larger number of atoms. This relationship is captured accurately in our VGOP calculations, with small particles clustered to the right and large particles clustered to left of Figure \ref{fig:C} (c). It is important to note here that the number of timesteps in each trajectory is not identical, so while the C70 bucky-ball appears to extend further right than the C60 particle, the C70 trajectory explores a significantly larger portion of its phase space than the C60 trajectory. 

\subsection{Aluminum}

The previous cases of lithium and carbon provided insight into the ability of the SGOP and VGOP frameworks to characterize both structural disorder and geometric diversity. For the case of aluminum, this coupling of complexity and heterogeneity is obtained by observing a multitude of non-zero temperature structural environments including surfaces, compressed and expanded lattices, point defects, grain boundaries, liquids, nanoparticles, all calculated previously via ab initio molecular dynamics \cite{Pun2019}. Here we compare our results to those of SP and the AGNI crystal fingerprint. SP represents a mathematically robust, though fixed with respect to any parameterization, characterization scheme that has been used to determine structural similarities for several decades. AGNI, on the other hand, represents a relatively new class of characterization schemes, in which structures are represented as a vector of highly parameterized functions, with each vector element capturing distinct parts of an atom's local geometric environment. By taking the PCA of SP, AGNI, and VGOP, we can create a level playing field, in which a direct comparison can be made between all three methodologies and their ability to characterize the same set of structural environments. Taking the PCA of feature sets has been used previously to visualize AGNI's ability to characterize atomic structures  \cite{doi:10.1021/acs.jpcc.9b03925}. 

\begin{figure*}
        \centering
    	\includegraphics[width=1.0\textwidth]{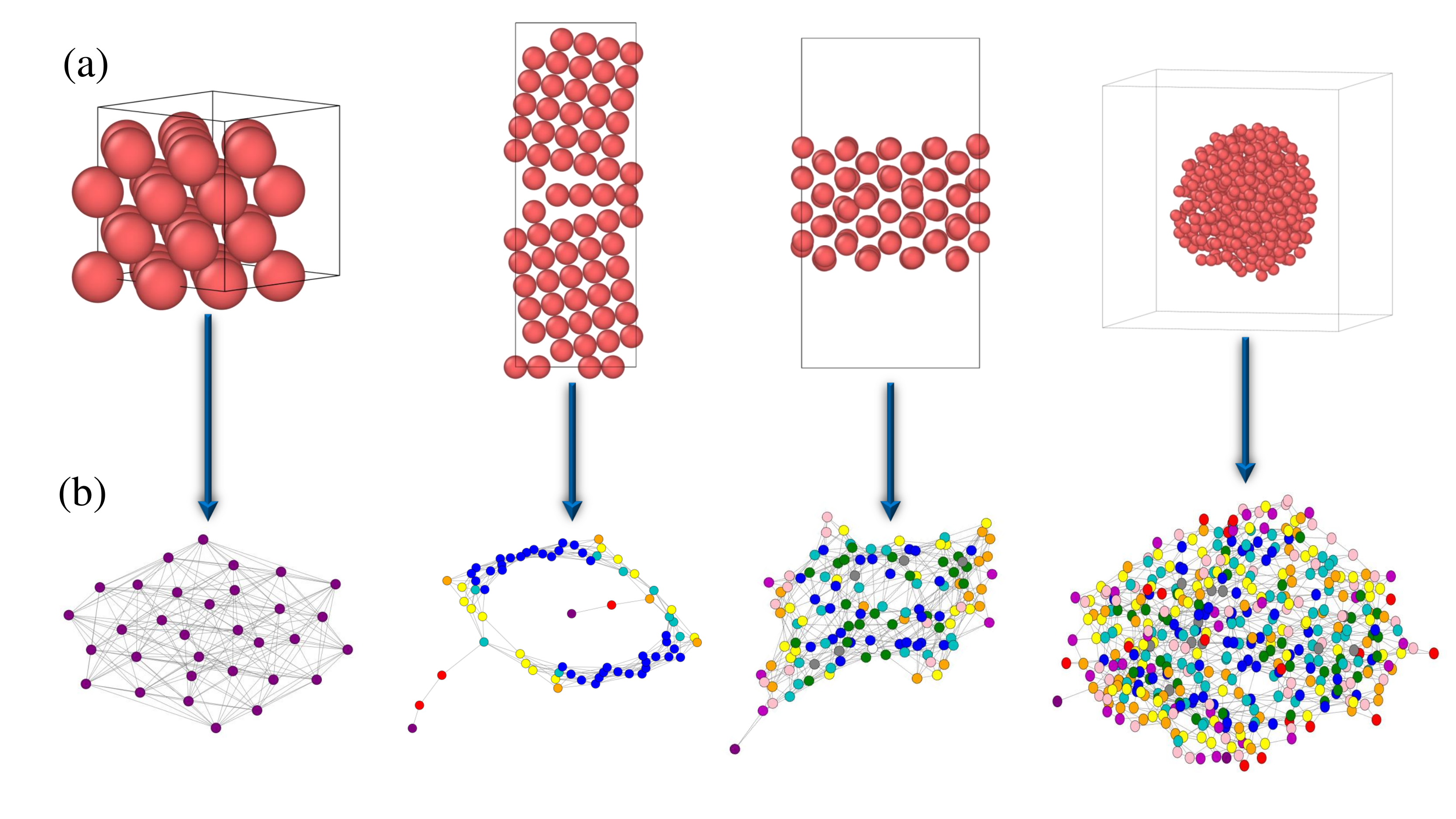}
    	\caption{(a) Aluminum atomic configurations from left to right: bulk FCC,  $\sum (510)$ grain boundary, $(110)$ surface, and a 12$\AA$ nanoparticle. (b) The corresponding graph coordination network at an $R_{c}$ of 3$\AA$. Vertices represent the atoms within their respective structures shown in (a), while the vertex colors represent the degree of the vertex. It should be noted that the colors are not universal, but are relative to the smallest vertex degree within the graph, with purple representing the smallest degree with blue being indicative of the largest degree.}
    	\label{fig:graphs}
\end{figure*}

We use the VGOP framework, with a $R_{c}$ set of $(3 $\AA$, 4 $\AA$, 5  $\AA$, 7  $\AA$, 8  $\AA$)$ determined via the aluminum RDF peaks. A visual representation of the GCN for $R_{c} = 3 \AA$ for several of the Aluminum structures is shown in Figure \ref{fig:graphs}. From Figure \ref{fig:graphs} one can see how the GCNs capture unique information about the structure. In the case of bulk Al the GCN indicates high but uniform connectivity amongst the nodes, while for the case of the grain boundary there exists two distinct regions of the graph, one corresponding to the bulk-like region and the other representing the interface region. A similar graph structure exists in the surface, though the surface region is far more chaotic and randomized than the fairly ordered grain boundary interface region. These structural differences within the graph provide a unique mapping from structure to VGOP, implying that structures with similar VGOP must have similar structural environments (provided one captures all relevant information via the cutoff radii).

\begin{figure*}
        \centering
    	\includegraphics[width=1.0\textwidth]{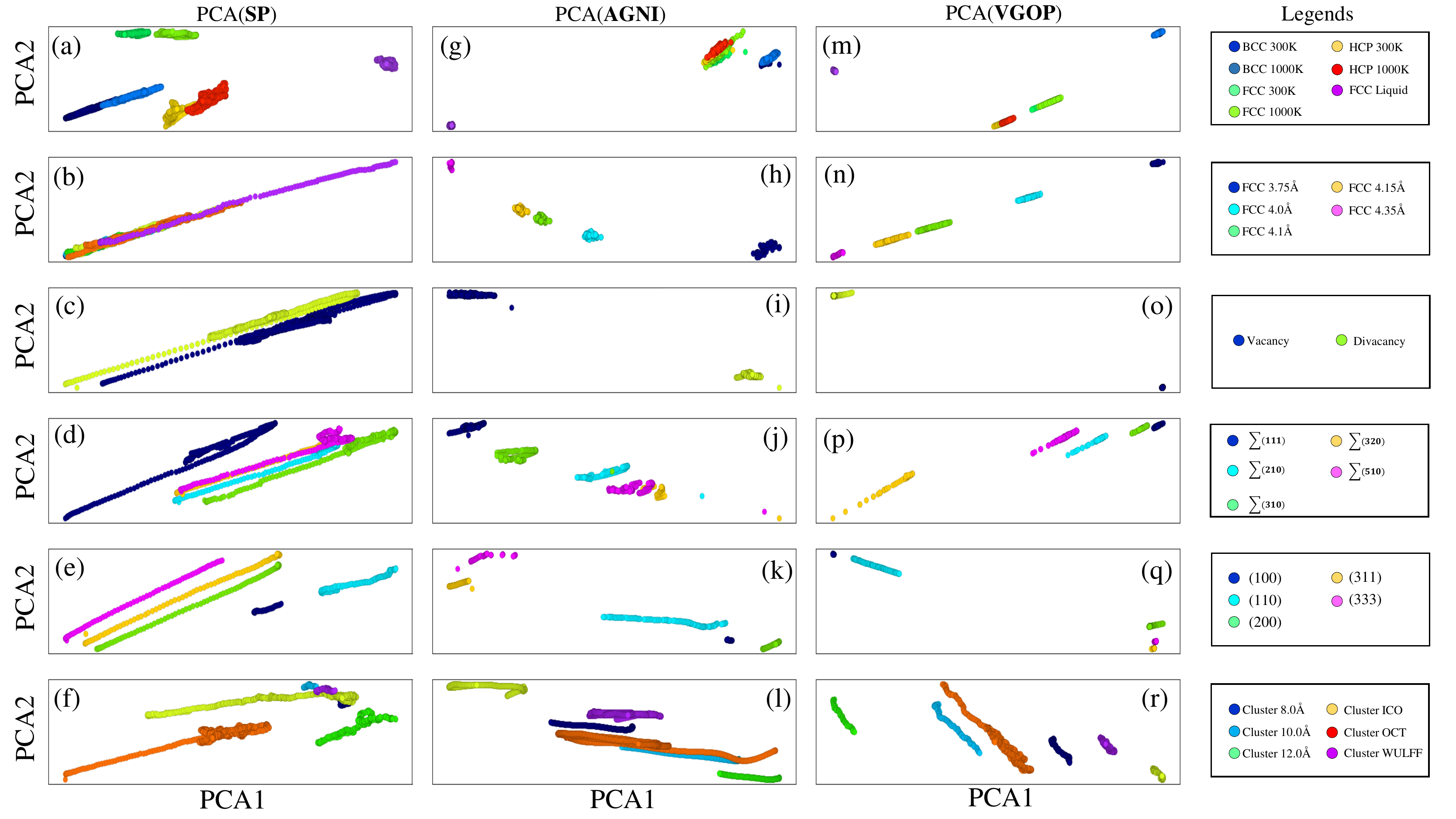}
    	\caption{Comparison of the PCA reduction of three different methods, using a robust aluminum database: (column 1) the Steinhardt order parameter, (column 2) the AGNI crystal fingerprint, and (column 3) the VGOP. Each row represents a unique subset of the aluminum configuration space. Colors are uniform across the columns. Each PCA subplot's axis have been normalized to the data present within the plot to allow for better visualization of the data. }
    	\label{fig:Al}
\end{figure*}

 For each method, a PCA decomposition was performed on the initial feature vector (i.e., computed SGOPs for each $R_{c}$ value), with the first two principal components chosen for visualization purposes. Such a procedure has been shown previously to be an accurate way of visualizing the high-dimensional spaces used in the structure characterization techniques used here \cite{doi:10.1021/acs.jpcc.9b03925}. Figure  \ref{fig:Al} showcases each method's ability to accurately characterize each class of aluminum environments. One should note here that each subplot's axes have been normalized between zero and one for visualization purposes, and that the absolute axis values between subplot are not shared. Further information regarding the details of each method can be found in the supplemental information.

The first column in Figure~\ref{fig:Al} represents the Steinhardt order parameter PCA classification. From Figure \ref{fig:Al} (a) one can see that the SP PCA eigenvectors can clearly distinguish between the low and high temperature BCC, FCC, and HCP phases. It also performs well when classifying the FCC liquid phase as distinct from the ordered crystal phases. However, the SP struggles to identify high temperature BCC as having the same underlying coordination environment as low temperature BCC. We know from the length of the trajectories that the high temperature structure has large thermal fluctuations of the ions that can mask its symmetry. In addition, Figure \ref{fig:Al} (b) indicates the SP's inability to correctly identify the structural differences between compressed and expanded FCC lattices, effectively characterizing all cases as a single entity. Figure \ref{fig:Al} (c) also highlights the SP's difficulty when attempting to differentiate between a single vacancy within a pristine bulk environment and that of a di-vacancy in an otherwise identical geometry. 

A similar trend emerges when characterizing the subtle differences in grain boundary structures, shown in Figure \ref{fig:Al} (d). The $\sum (210)$, $\sum (310)$, and $\sum (510)$ grain boundaries should yield some underlying similarities, but are technically unique environments. However, the SP has difficulty in distinguishing between the configurations, and also classifies the $\sum (510)$ and $\sum (320)$ grain boundaries as identical coordination environments, which is incorrect. Interestingly the SP performs well when characterizing the differences between surface environments in Figure \ref{fig:Al} (e), perhaps due to the well-defined uniqueness in the surface layers. For the case of the nanoparticles, shown in Figure \ref{fig:Al} (f), the SP is able to clearly differentiate between the ordered clusters (iscohedral, octohedral, and Wullf particles), but fails to correctly capture the differences inherent in the disordered particles (8.0 $\AA$, 10.0 $\AA$, and 12.0 $\AA$ particles). All told, the SP cannot be reliably used to characterize the complexity of the aluminum configuration space.

The second column represents the AGNI crystal fingerprint classification. From Figure \ref{fig:Al} (g) one can see that the AGNI PCA eigenvectors can clearly distinguish between the low and high temperature BCC and HCP/FCC, but fails to correctly capture the differences between HCP and FCC. However, it does perform well when classifying the FCC liquid phase as distinct from the ordered bulk phases. Figure 4h indicates AGNI's ability to correctly identify the structural differences between compressed and expanded FCC lattices. Figure \ref{fig:Al} (i) highlights AGNI's capabilities in differentiating between the vacancy and divacancy environments. Unlike the SP, AGNI performs much better when characterizing the subtle differences between grain boundaries, though does encounter some overlap between the $\sum (510)$ and $\sum (320)$ structures. AGNI also performs well when characterizing the differences between surface environments in Figure \ref{fig:Al} (k). However, for the case of the nanoparticles, shown in Figure \ref{fig:Al} (l), AGNI fails to properly distinguish between the ordered and disordered clusters, similar to the problematic characterization of the SP. Overall, while AGNI can correctly capture a much larger portion of the aluminum configuration space, it breaks down in several areas, some of which could be correctly captured by the SP.

The third column represents the VGOP classification. From observing Figures \ref{fig:Al} (m)-(r) one can see that the VOP framework predicts a unique characterization for every structural environment encountered in the dataset. Perhaps equally as important is the VGOP's ability to cluster similar coordination environments together, providing an intuitive and natural unsupervised clustering. In princial, if one did not know what the structures being characterized were, they could identify geometric similarites, or differences, between them. Having this ability could make the VGOP a powerful tool for enhancing sampling methods during model development. One could use the VGOP to indicate structures that a model does not need to be parameterized on, due to the underlying similarities with other environments. 

\section{Discussion}

Structure-property relationships, which have always served a fundamental role in materials science, have become critically important due to the ever-increasing need for new materials with targeted properties and chemistries. Frequently, a high degree of precision and accuracy can be needed to uniquely characterize new structural environments and ultimately map those unique geometries to properties of interest. Our efforts here discusses a new structural  characterization scheme that can uniquely classify the local atomic coordination environments present in atomistic configurations through a semi-emprical graph isomorphism order parameter. Our formalism is computationally efficient and mathematically robust, providing the ability to characterize subtle differences in atomic structure over a wide range of thermodynamic conditions. While the SGOP formalism requires minimal user-adjusted parameters (such as the graph $R_{c}$ and SGOP exponent), they are physically intuitive, and require only a limited understanding of the underlying system to be appropriately chosen.

Contrary to many popular machine learning methodologies such as CNNs, GNNs, and VAEs, the SGOP/VGOP formalism is an absolute metric that, once a set of parameters has been chosen, can be used arbitrarily for any material system. Finally, the computational cost of the SGOP/VGOP framework is minimal compared to other methods such as SOAP, BP, and AGNI, as the only loop contained within the mathematical formalism is with respect to the graph's degree matrix. Our algorithm is also easily parallelizable, and can be determined efficiently on any modern computing system. The computational efficiency combined with the uniqueness and physically-informed nature of the formalism allows it to be applied to a plethora of challenging application spaces including enhanced sampling, and unsupervised clustering, which generally require the ability to determine subtle distinctions between underlying phases or structures.

\section{Computational Details}
\subsection{Graph Coordination Networks}

The diversity and complexity of atomic structures neccesitates the efficient and intuitive characterization of these evironments. In this work, we employ a graph-based characterization scheme, which we call the Graph Coordination Network, to identify pairwise atomic networks contained within a configuration of atoms. GCNs begin by sorting the chemical identities of the atoms in the configuration into separate categories. Depending on the pairwise interaction one aims to capture, the various species lists are then scanned to find atomic interactions that occur within some cutoff radius. The GCN is similar to a radial distribution function, as it aims to capture the unique coordination environments encountered by each atom, with respect to a particular chemical interaction environment. The GCN can be represented by an adjacency matrix, with matrix elements defined by:

\begin{equation}
	G_{k_{i},k_{j}}^{i-j} = \frac{1}{d_{k_{i},k_{j}}}  \ni d_{k_{i},k_{j}} \leq R_{c}
\label{equ:GCN}
\end{equation}

 Here, $i$ and $j$ represent the chemical identities of the atoms contained in the GCN. $k_{i}$ and $k_{j}$ are the atomic index of a given atom from chemical specie $i$ and $j$ respectively. $d_{k_{i},k_{j}}$ is defined as the $l^{2}$-norm between two atoms. $R_{c}$ is the cutoff radius specified when constructing the GCN. A visual depiction of how a GCN is constructed from various Aluminum atomic structures can be found in Figure \ref{fig:graphs}. Each matrix element, $\frac{1}{d_{k_{i},k_{j}}}$, represents the weight of a given edge for a given pair of connected nodes in the graph. The degree of each node is then given by the sum of the elements in a node's edge set. It should be noted that the matrix representation of the GCN is equivalent to a Coulomb matrix \cite{Schrier2020,https://doi.org/10.1002/qua.24917}, which has been used previously to characterize molecular environments. 

\subsection{Scalar Graph Order Parameter}

 Here, we introduce the scalar graph order parameter to characterize the atomic coordination networks contained within the GCNs described in the previous section. Generally speaking, one can think of SGOP as a semi-empirical physically-informed graph similarity metric. We define this SGOP as:

\begin{equation}
    \theta_{i-j,R_{c}} =  \sum_{s}^{S}\left( \sum_{m}^{D_{s}}P(d_{m})\log_{b}P(d_{m}) + d_{m}P(d_{m})  \right)^{3}
\label{equ:SGOP}
\end{equation}

Here, $i$ and $j$ represent the chemical identities of the atoms contained in the GCN. $R_{c}$ is the cutoff radius specified when constructing the GCN. We make the assumption that a particular GCN is disconnected, and that the underlying network exists as a set of subgraphs, $S$, with $s$ indexing a particular subgraph. Note that in the event a GCN is fully connected the outer sum dissappears and no further changes are required to the formalism. $D_{s}$ is the set of unique node degrees in a subgraph, with $P_{d_{m}}$ being the probability of a given degree, $d_{m}$, occuring in the subgraph.

The underlying formalism of SGOP provides physical intuition about a graph: (a) $P(d_{m})\log_{b}P(d_{m})$ uses entropy to approximate a graph's shape, and (b) $d_{m}P(d_{m})$ characterizes a graph's connectivity. The entropic term can be easily identified as capturing the amount of ``chaos'' present in a graph, providing a unique mapping to the underlying shape. The connectivity term represents an empirical approximation of the ``density'' of a graph. It can be insufficient to compare more standard graph properties such as the maximum degree, minimum degree, and average degree, as these metrics can be not unique enough to capture the diversity present in a material's phase space. Therefore, the connectivity term was crafted to identify not only the degrees present within a graph but also the likelihood of occurrence of those degrees.  

The cubed exponent of the inner summation provides a heuristic weighting mechanism to compare the sum of entropy and connectivity that was determined through trial and error. It is important to remember that the SGOP value is simply the sum of subgraph SGOPs if multiple subgraphs are present within a configuration of atoms. If the exponent is too large, highly connected and chaotic subgraphs will always be weighted too heavily when compared to smaller, poorly connected subgraphs. If the exponent is too small the opposite becomes true, in which subgraphs that are explicitly distinct run the risk of becoming indistinguishable during unsupervised clustering. Our experimentation has indicated that a cubed exponent provides the strong balance between these two extreme scenarios. In this way, SGOP can capture both similarities and subtle differences between graphs in a computationally efficient manner. 

We note that the SGOP formalism is generalizable and transferable to any any graph characterization, and is not restricted to the study of atomic configurations. It should also be noted that equation \ref{equ:SGOP} is invariant under permutation, translation, replication (system size), and rotation operations. We also note that there exist a multitude of graph-based formalisms in the literature that aim to characterize atomic structures \cite{doi:10.1021/acs.jcim.9b00410,WODO20121105,ESTRADA2000713,hall_electrotopological_1991}, and the primary distinction between such methods and those prescribed in this work is the computational cost, mathematical simplicity, and universal transferability of our method. While further details regarding the software formalism and cost of the SGOP calculations can be found in the supplementary information, we will indicate here that an SGOP for a 32,000 atom Aluminum system was computed in less than 0.5 seconds using only a serial execution. The low-cost of the algorithm allows for the efficient characterization of not only complex structural systems, but also the study of systems on the order of tens of nanometers in size.

\begin{figure*}
        \centering
    	\includegraphics[width=1.0\textwidth]{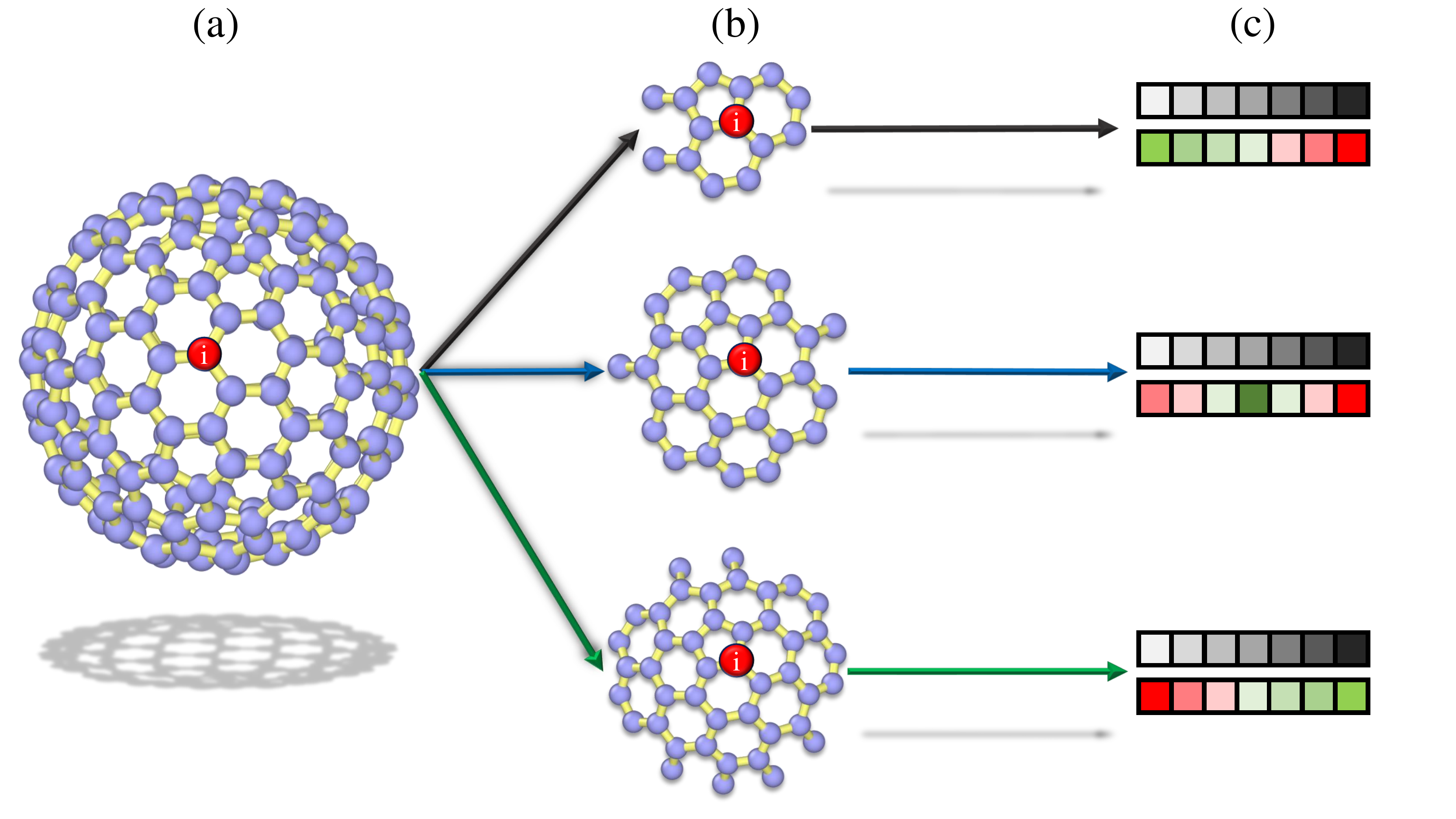}
    	\caption{Visualization of the VGOP framework. (a) A 180-atom buckyball is shown, with a specific atom $i$ highlighted in red. (b) 3 substructures of the buckyball, determined by selecting atoms varying cutoff radii away from atom $i$, that form the basis of unique GCNs. (c) 3 sets of 2 bars, which each subset representing the degree set (top) of each GCN and the degree probability set (bottom). The colors of the degree set are defined as white representing low connectivity and black representing high connectivity. The colors of the probability set are specified as green representing high likelihood and red representing low liklihood of occurance in the GCN. The varying colors of the probability sets indicate that the smallest substructure in (b) yields a large number of poorly connected nodes and little-to-no high connectivity, while the opposite is true for the largest substructure in (b).  }
    	\label{fig:VGOP}
\end{figure*}

\subsection{Vector Graph Order Parameter}

While the SGOP formalism prescribed in the previous section accurately characterizes the graph network encoded within an atomic environment, the resulting value encodes local geometric information within a coordination sphere of radius $R_{c}$. Many atomic environments share underlying similarities in their local structure, which leads to overlapping values within the order parameter space. As a result, a single scalar is often not sufficient to distinguish between the complexity of a material's configuration space due to seemingly small but important differences encountered between atomic systems.

Here, we introduce the Vector Graph Order Parameter, which is simply a set of SGOP values, calculated using a unique, user-chosen set of $R_{c}$. By taking a set of coordination sphere radii, one can ensure that various portions of an atom's local geometry are properly encoded. Figure \ref{fig:VGOP} shows a visual workflow for how the VGOP is determined for the case of a carbon nanoparticle. Principle component analysis \cite{karamizadeh_overview_2013} is used to reduce the number of features and allow for the visual inspection of the underlying data. Z-score normalization \cite{8667324} was used to normalize the VGOPs as a preprocessing measure to aid in the PCA decomposition, though in principle is not necessary. For the material systems studied in this work the first two principle components comprised at least 95$\%$ of the underlying variance, and therefore the remaining components were discarded. Further information regarding the PCA decomposition for all systems studied in this work can be found in the supplementary information.

\section*{Acknowledgements}
We would like to thank Sabri Elatresh and Stanimir Bonev for allowing us to use their DFT liquid lithium database. J. Chapman, N. Goldman, and B. Wood are partially supported by the Laboratory Directed Research and Development (LDRD) program (20-SI-004) at Lawrence Livermore National Laboratory. This work was performed under the auspices of the US Department of Energy by Lawrence Livermore National Laboratory under contract No. DE-AC52-07NA27344.

\begin{suppinfo}
The supporting information for this work can be found online.
\end{suppinfo}
\providecommand{\noopsort}[1]{}\providecommand{\singleletter}[1]{#1}%
\providecommand{\latin}[1]{#1}
\makeatletter
\providecommand{\doi}
  {\begingroup\let\do\@makeother\dospecials
  \catcode`\{=1 \catcode`\}=2 \doi@aux}
\providecommand{\doi@aux}[1]{\endgroup\texttt{#1}}
\makeatother
\providecommand*\mcitethebibliography{\thebibliography}
\csname @ifundefined\endcsname{endmcitethebibliography}
  {\let\endmcitethebibliography\endthebibliography}{}

\end{document}